\documentclass[sn-mathphys,Numbered]{sn-jnl}
\usepackage{graphicx}%
\usepackage{multirow}%
\usepackage{amsmath,amssymb,amsfonts}%
\usepackage[T1]{fontenc}
\usepackage{amsthm}%
\usepackage{mathrsfs}%
\usepackage[title]{appendix}%
\usepackage{xcolor}%
\usepackage{textcomp}%
\usepackage{manyfoot}%
\usepackage{booktabs}%
\usepackage{algorithm}%
\usepackage{algorithmicx}%
\usepackage{algpseudocode}%
\usepackage{listings}%
\usepackage{float}
\theoremstyle{thmstyleone}%
\theoremstyle{thmstyletwo}%
\theoremstyle{thmstylethree}%
\begin{document}
\title[Article Title]{Noninvasive megapixel fluorescence microscopy through scattering layers by a virtual reflection-matrix}

\author[1]{\fnm{Gil} \sur{Weinberg}}
\equalcont{These authors contributed equally to this work.}

\author[1]{\fnm{Elad} \sur{Sunray}}
\equalcont{These authors contributed equally to this work.}

\author*[1]{\fnm{Ori} \sur{Katz}}\email{orik@mail.huji.ac.il}

\affil[1]{\orgdiv{Institute of Applied Physics}, \orgname{The Hebrew University of Jerusalem}, \orgaddress{\city{Jerusalem}, \postcode{9190401}, \country{Israel}}}

\maketitle

\section*{Abstract}

Optical-resolution fluorescence imaging through and within complex samples presents a significant challenge due to random light scattering, with substantial implications across multiple fields. 
While significant advancements in coherent imaging through severe multiple scattering have been recently introduced by reflection-matrix processing, approaches that tackle scattering in incoherent fluorescence imaging have been limited to sparse targets, require high-resolution control of the illumination or detection wavefronts, or a very large number of measurements.
Here, we present an approach that allows direct application of well-established reflection-matrix techniques to scattering compensation in incoherent fluorescence imaging. 
We experimentally demonstrate that a small number of conventional widefield fluorescence-microscope images acquired under unknown random illuminations can effectively construct a fluorescence-based virtual reflection matrix. This matrix, when processed by conventional matrix-based scattering compensation algorithms, allows reconstructing megapixel-scale fluorescence images, without requiring the use of spatial-light modulators (SLMs) or computationally-intensive processing.

\section*{Introduction}

Fluorescence imaging is an essential tool in noninvasive biomedical research. Yet, its application for imaging in complex samples is challenging due to random light scattering, limiting the application of optical microscopes. Inherent scattering of light in biological specimens significantly reduces the sharpness and clarity of images, making it extremely difficult to obtain optical-resolution fluorescence images noninvasively \cite{yoon2020deep,bertolotti2022imaging}.
Recent advancements to overcome this challenge include speckle-correlation imaging \cite{bert12,katz14}, which relies on the inherent isoplanatism of multiple scattering in a limited angular range, commonly referred to as the optical memory effect \cite{feng1988correlations}.
Notable examples include computational approaches that retrieve the target object from the captured-image autocorrelation, as done in stellar speckle interferometry \cite{labeyrie1970attainment}. However, they require capturing a number of speckle grains that is considerably larger than the number of object resolution cells and rely on iterative phase retrieval, which lacks guaranteed convergence for complex targets. 
The reconstruction convergence challenge can be overcome by deterministic bispectrum reconstruction \cite{wu2016single}, leveraging the principle of closure phase from radio-astronomy but still requiring a very large number of speckle grains to be captured and computationally intensive processing.

A promising path for noninvasive imaging through complex media is made possible by coherent -based techniques that are based on phase-sensitive measurements of scattered coherent fields \cite{kang17,badon2020distortion,lee22,najar23,kang2023tracing,zhang2023deep}. In these works, computational phase correction allows the undoing of scattering and reconstruction of scattering-free images of complex reflective targets.
While these phase-sensitive methods have proven effective for coherent reflective imaging \cite{kang2023tracing,haim2023image}, they are incompatible with fluorescence imaging since the phase of fluorescence signals from extended objects is undefined.

Several attempts have been made to utilize matrix-based methods in the incoherent case, each with distinct challenges and requirements. Several techniques apply physical correction of either the excitation or detection path, or both, utilizing spatial light modulators (SLM). 
These approaches rely on measuring the feedback from different excitation patterns, for example, by maximizing the image variance \cite{daniel19,boniface19} or applying an incoherent iterative phase conjugation \cite{aizik2022fluorescent,aizik2023non,baek2023phase}. 
Correction of the detection path can be obtained by improving the image quality either iteratively \cite{yeminy2021guidestar}, or using a neural-networks to find the optimal correction \cite{feng23, d2022physics}. Matrix decomposition methods, particularly non-negative matrix factorization (NMF), while computationally very demanding, are helpful when the target is spatially very sparse, as they demand the number of acquired frames that is significantly larger than the number of bright resolution cells \cite{boniface20,zhu22}. 

Here, we introduce a novel framework that allows to computationally undo the effects of severe scattering in any conventional widefield fluorescence microscope without requiring an SLM, target sparsity, or a large number of captured frames. 
Our approach is based on constructing a matrix analog of the coherent from a few tens of fluorescence camera frames obtained under unknown random illuminations. This '', obtained through a straightforward cross-correlation computation, lends itself to processing by any of the well-established matrix-based scattering compensation schemes \cite{kang17,badon2020distortion}, as we experimentally demonstrate on isoplanatic-scattering samples. 
Specifically, by treating each camera pixel as a random variable and calculating the covariance matrix between camera pixel intensities over random unknown illuminations, a reflection-like matrix with dimensions of $N \times N$ where $N>10^6$ is the number of camera pixels is formed. Importantly, we demonstrate that the matrix can be formed even when the number of measurements, $M$, is smaller by more than four orders of magnitude than the number of camera pixels.
This covariance analysis resembles Green's function retrieval process in passive correlation imaging investigated in seismology and acoustics \cite{derode2003recovering,wapenaar2004retrieving,garnier2016passive}, and the compressed time-reversal (CTR) reflection-matrix acquisition scheme \cite{lee22}, but here for the first time for spatially-incoherent signals.

To efficiently process the reflection matrix, which contains over a trillion ($N^2 > 10^{12}$) elements, and address the challenge of amplitude modulation in incoherent scattering, we propose a mathematically equivalent yet memory-efficient algorithm to CTR-CLASS (closed-loop accumulation of single scattering). This algorithm necessitates the storage of only $O(MN)$ elements. Additionally, it includes an extra post-processing step to correct distortions in the Modulation Transfer Function ($MTF = |OTF|$).
We experimentally demonstrate reconstructions of $>10^6$ pixels widefield fluorescence images of non-sparse targets, correcting more than $10^4$ k-space modes, using only $150$ acquired camera frames. 

\section*{Results}
\subsection*{Principle}

\begin{figure}[htb!]
        \vspace{-20pt}
	\centering
	\includegraphics [width=\textwidth,]
	{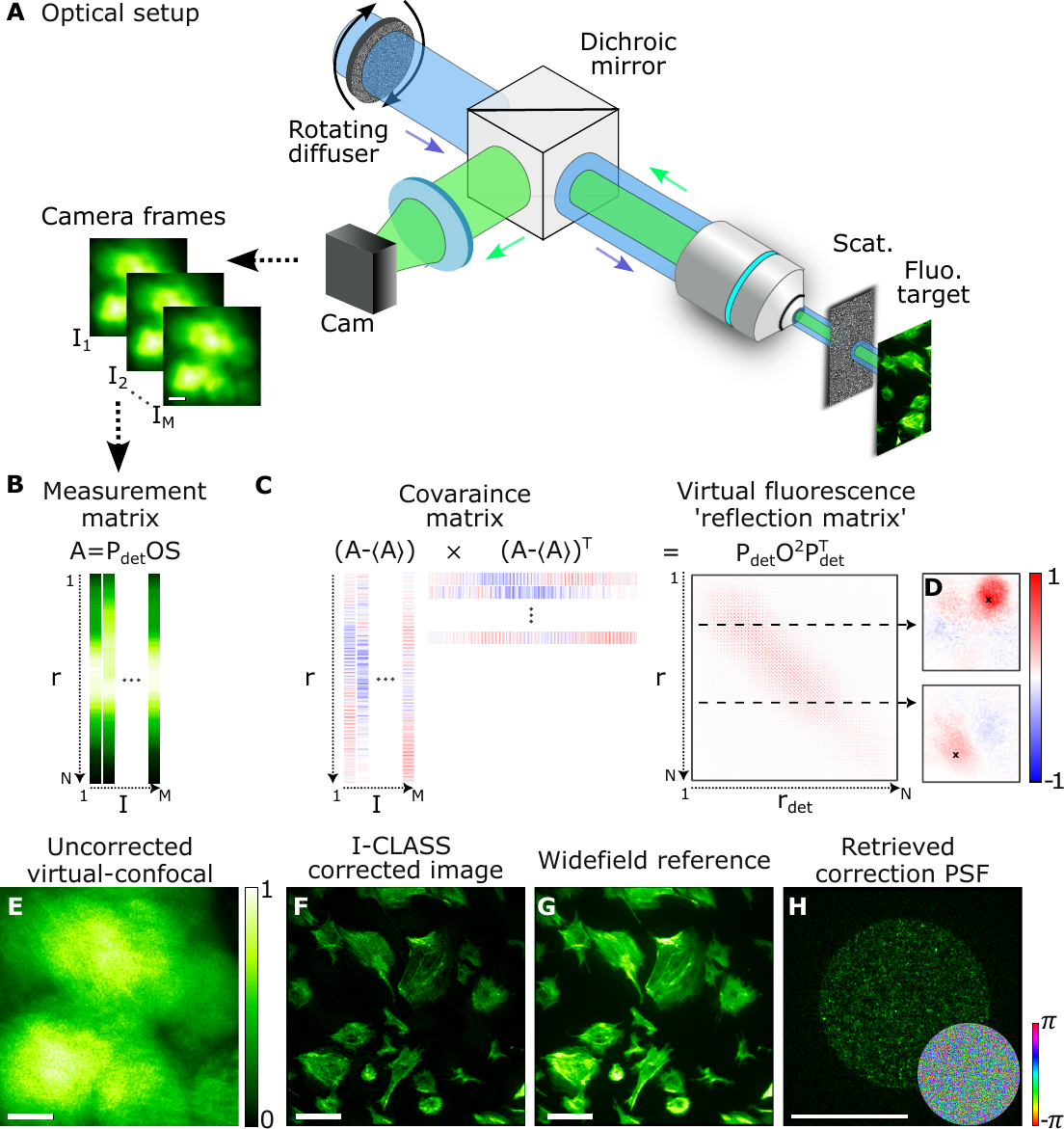}
	\caption{\textbf{Noninvasive imaging through scattering layers with a fluorescence virtual reflection-matrix, concept and numerical example}
 (A) Setup: A fluorescence target hidden behind a highly scattering layer is imaged by a conventional widefield fluorescence microscope under random unknown speckle illuminations. 
 The $M$ recorded camera frames, $I_1(\vec{r}),...,I_M(\vec{r})$, are reshaped into the columns of a measurement matrix, $\textbf{A}$.
 (B) The measurement matrix $\textbf{A}$ is given by: $\textbf{A}=\textbf{P}_{det}\textbf{O}\textbf{S}$, where $\textbf{S}$, $\textbf{O}$, and $\textbf{P}_{det}$ denote the unknown speckle illumination patterns, the object fluorescence labeling distribution, and the highly-distorted diffusive scattering PSF, respectively. 
 (C) Since ${\hat{\textbf{S}}} \hat{\textbf{S}}^T \propto \textbf{I}$ the covariance matrix of $\textbf{A}$ has the form of a virtual 'reflection matrix': $\textbf{P}_{det}\textbf{O}^2\textbf{P}_{det}^T$, where each of its columns (or rows) represents a widefield image that would be recorded when the illumination is virtually focused to a point, $\vec{r}_n$, on the target (D, marked by small 'x's). The virtual 'reflection matrix' diagonal provides a virtual confocal image (E).
(F) Applying a CLASS reflection-matrix-based scattering-compensation algorithm, modified for the spatially incoherent measurements, provides a corrected image of the target (F), as would be obtained without the scattering layer (G). (H) The modified CLASS algorithm (I-CLASS) also provides the scattering PSF, and the k-space correction phases (inset). 
 Scale bars, 100 pixels.
    }
	\label{fig1}
\end{figure}

The principle of our method, together with a numerically simulated example (see Methods), is schematically shown in Fig.~\ref{fig1}. Our approach for imaging a fluorescence target behind a highly scattering layer is based on a conventional widefield fluorescence microscope that images the fluorescent sample under several random unknown illuminations (Fig.~\ref{fig1}A). In the presented implementation, the random illumination is provided in epi-configuration by passing a continuous wave (CW) laser (Fig.~\ref{fig1}A, in blue) through a rotating diffuser. 
An sCMOS camera captures $M$ images of the scattered fluorescence light through a dichroic mirror and detection filters in a conventional $4f$ imaging setup. The captured frames are thus identical to those obtained by a dynamic speckle illumination microscope \cite{ventalon2005quasi, mudry2012structured}.

Following the acquisition process, we form a virtual 'reflection matrix' from the covariance between the different camera pixels (Fig.~\ref{fig1}B, C). This covariance matrix can be straightforwardly computed from the raw low-contrast camera frames by first forming an $N\times M$ 'measurement matrix', $\textbf{A}$, where each of its $M$ columns is a single raw camera frame (Fig.~\ref{fig1}B). 
As we show below, the covariance matrix, $\textbf{R}$, of this measurement matrix $\textbf{A}$, which is simply the cross-correlation of $\textbf{A}$, after subtraction of its row-wise mean $\vec{\langle \textbf{A}\rangle}$: $\textbf{R}=(\textbf{A}-\vec{\langle \textbf{A}\rangle})(\textbf{A}-\vec{\langle \textbf{A}\rangle})^T$, can be interpreted as a virtual 'reflection matrix' for fluorescence. In this virtual reflection matrix, each camera pixel serves both as a detector and a virtual illumination point source. Each column (or row) of $\textbf{R}$ thus provides the camera image that will be captured when a virtual illumination originates from each single camera pixel (Fig.~\ref{fig1}D).
  
As in a conventional reflection matrix, the virtual 'reflection matrix' diagonal provides a 'confocal' image (Fig.~\ref{fig1}E). While this confocal-like image, which is equivalent to the variance analysis of dynamic speckle illumination microscopy \cite{ventalon2005quasi}, or speckle-SOFI \cite{dertinger2009sofi, kim2015superresolution}, is sharper and has higher contrast than the raw camera frames, it is still severely distorted and blurred by scattering. High-resolution scattering compensation must be employed to reconstruct the hidden fluorescent target. 
Elegantly enough, as we show below, the virtual 'reflection matrix', $\textbf{R}$ lends itself to analysis and processing by any of the well-established scattering-correction techniques developed for the conventional coherent complex-valued reflection matrix \cite{kang17,badon2020distortion}.
As a demonstration, we apply a modified-CLASS algorithm \cite{kang17,lee22} (see Fig.~\ref{fig2}) to effectively correct the effects of isoplanatic scattering in this numerical demonstration.
Indeed, the CLASS algorithm provides both the corrected image (Fig.~\ref{fig1}F), which restores the high-resolution fine features of the target (Fig.~\ref{fig1}G), and an estimation of the scattering PSF (Fig.~\ref{fig1}H). 

\subsection*{Covariance matrix equivalence to a virtual 'reflection matrix'}

Here, we establish the mathematical foundation supporting the preceding claims.
Assuming isoplanatism, the relationship between each camera frame, $I_m(\vec{r})$, under the $m=1...M$ random speckle illumination, $S_m(\vec{r})$, and the fluorophores distribution of the target object, $O(\vec{r})$, is given by a convolution with the detection point-spread-function (PSF) $P_{det}(\vec{r})$: 

\begin{eqnarray}
I_{m}(\vec{r})=P_{det}(\vec{r})*[O(\vec{r})\cdot S_{m}(\vec{r})]
\label{eq:one}
\end{eqnarray} 

\noindent By arranging the captured camera frames into columns of a 'measurement matrix' $\textbf{A}$ (Fig.~\ref{fig1}B), we can write Eq.~\ref{eq:one} in a similar way to \cite{lee22}:
\begin{eqnarray}
    \textbf{A} = \textbf{P}_{det}\textbf{O}\textbf{S}
    \label{eq:onematrix}
\end{eqnarray}
Where $\textbf{S}$ is a matrix with each of its columns containing the random (unknown) speckle illumination patterns at the object plane, $\textbf{O}$ takes the form of a diagonal matrix with the object fluorescence distribution $O(\vec{r})$ as its diagonal elements, and $\textbf{P}_{det}$ is a convolution matrix representing the (unknown) randomly distorted detection PSF. In the case of imaging through highly scattering layers, the PSF is a random unknown speckle intensity pattern. The goal of a noninvasive imaging technique is to retrieve \textbf{O} (and potentially also $\textbf{P}_{det}$) from \textbf{A}, where $\textbf{P}_{det}$,\textbf{O} and \textbf{S} are unknown matrices.

Given that $\textbf{S}$ consists of uncorrelated speckle-intensity patterns, the covariance matrix of $\textbf{S}$ can be approximated by the identity matrix $\textbf{I}$. This covariance matrix can be calculated by multiplying $\textbf{S}$ by its transpose after performing a row-wise mean reduction:

${\hat{\textbf{S}}} \hat{\textbf{S}}^T \propto \textbf{I}$,
where $\hat{\textbf{S}}\stackrel{\text{def}} = \textbf{S} (\mathbf{I}-\frac{1}{M}\overline{\textbf{1}})$ is the mean subtracted speckle illumination matrix ($\overline{\textbf{1}}$ is a matrix comprising entirely of ones, see supplementary S1).

Notably, using the decomposition of Eq.~\ref{eq:onematrix}, it becomes apparent that subtracting the row mean from the known measurement matrix $\textbf{A}$ – that is, removing the temporal mean value from each camera pixel across the speckle illuminations – is equivalent to subtracting the row mean from the unknown illumination patterns:

\begin{eqnarray}
\hat{\textbf{A}}\stackrel{\text{def}} = \textbf{A} - \vec{\langle \textbf{A}\rangle} \equiv \textbf{A}(\mathbf{I}-\frac{1}{M}\cdot\overline{\textbf{1}}) = {\textbf{P}_{det}}{\textbf{O}}{\textbf{S}}(\mathbf{I}-\frac{1}{M}\cdot\overline{\textbf{1}})
 = \textbf{P}_{det}\textbf{O}\hat{\textbf{S}}
\end{eqnarray}
As a direct result, calculating the covariance matrix of the matrix \textbf{A}, provides a fluorescence virtual 'reflection matrix', where both the input and output distortions are given by the distorted detection PSF, and the random speckle illuminations play no role: 

\begin{eqnarray}
\textbf{R}=\hat{\textbf{A}}\hat{\textbf{A}}^T = \textbf{P}_{det}\textbf{O}\hat{\textbf{S}}\hat{\textbf{S}}^T\textbf{O}\textbf{P}_{det}^T \approx {\textbf{P}_{det}}{\textbf{O}_{eff}} {\textbf{P}_{det}^T} \label{eq:two}
\end{eqnarray} 
Where $\textbf{O}_{eff}\stackrel{\text{def}}=\textbf{O}^2$, in a similar fashion to the coherent case of CTR-CLASS \cite{lee22}.

The reason why $\textbf{R}$ can be treated as a virtual 'reflection matrix' originates from the analogy to the conventional coherent reflection matrices \cite{kang17,lee22,badon2020distortion}: In an isoplanatic coherent optical system, the coherent reflection matrix is given by: $\textbf{R}_{coh}=\textbf{P}_{det}\textbf{O}_{coh}\textbf{P}_{ill}^T$ where $\textbf{P}_{det}$ and $\textbf{P}_{ill}$ are complex-valued convolution matrices representing the convolution with the detection and illumination coherent PSFs, respectively, and $\textbf{O}_{coh}$ is a diagonal matrix with the object's reflectivity function on its diagonal.
This equivalence in mathematical formulation thus allows for applying the well-established algorithms designed for coherent matrix imaging in the fluorescence imaging case. Specifically, one can apply the state-of-the-art scattering compensation algorithms of CLASS \cite{kang17} and distortion-matrix \cite{badon2020distortion}. We demonstrate the effectiveness of such processing by applying a modified version of the CLASS algorithm to incoherent fluorescence imaging.

\subsection*{Modified and memory-efficient CLASS}
The standard CLASS algorithm corrects the effects of scattering in the coherent imaging case by applying a phase-only correction in the Fourier domain of the object plane \cite{kang17}. CLASS has been proven extremely effective in undoing aberrations and scattering of tissue \cite{kang17}, mouse skull \cite{kwon2023computational}, and multicore fibers \cite{choi2022flexible}, both in reflection imaging and second-harmonic generation (SHG) microscopy \cite{moon2023synthetic, murray2023aberration}. The correction phase is found from correlations between the coherent reflection matrix columns (see Supplementary section S1).
In incoherent imaging, isoplanatic scattering is manifested as a multiplication by a complex-valued distorted optical transfer function (OTF), which is the Fourier transform of the incoherent PSF. Since the OTF is the autocorrelation of the complex pupil function \cite{goodman2005introduction}, in the case of random scattering, strong amplitude modulations are present within the OTF. Thus, applying the phase-only standard CLASS correction still results in a residual significant 'haze' in the reconstructed image (Fig.~\ref{fig2}). To mitigate this, we have extended the CLASS algorithm to estimate not only the k-space phase distortions but also the amplitude modulations. This is achieved by utilizing the virtual 'reflection matrix' in the Fourier domain (see Methods). The phase correction is applied as in conventional CLASS, and the amplitude compensation is applied by a regularized Fourier reweighting (see Methods and Supplementary Section 1). We term this modification I-CLASS (Incoherent-CLASS).

While in principle, one can apply CLASS on the $N \times N$ matrix $\textbf{R}$, where $N$ is the number of camera pixels, in practice, when considering megapixel scale images, the resulting 'virtual reflection matrix' $\textbf{R}$ would require more than 1TB of memory. To allow CLASS processing of such megapixel scale images, we have developed an alternative method to perform the CLASS iterations directly on the matrix $\hat{\textbf{A}}$, avoiding the direct computation of $\textbf{R}$ (see supplementary S1). Since $\hat{\textbf{A}}$ is of dimensions $N \times M$, where  $M\approx150\ll N$, this approach significantly reduces the memory requirements to $O(MN)$ from the previously required $O(N^2)$, thereby enabling the algorithm to run on over $10^6$ pixels from hundreds of images using standard computer hardware. 

\subsection*{Experimental results}

\begin{figure}[H]
	\centering
	\includegraphics [width=\textwidth,]
	{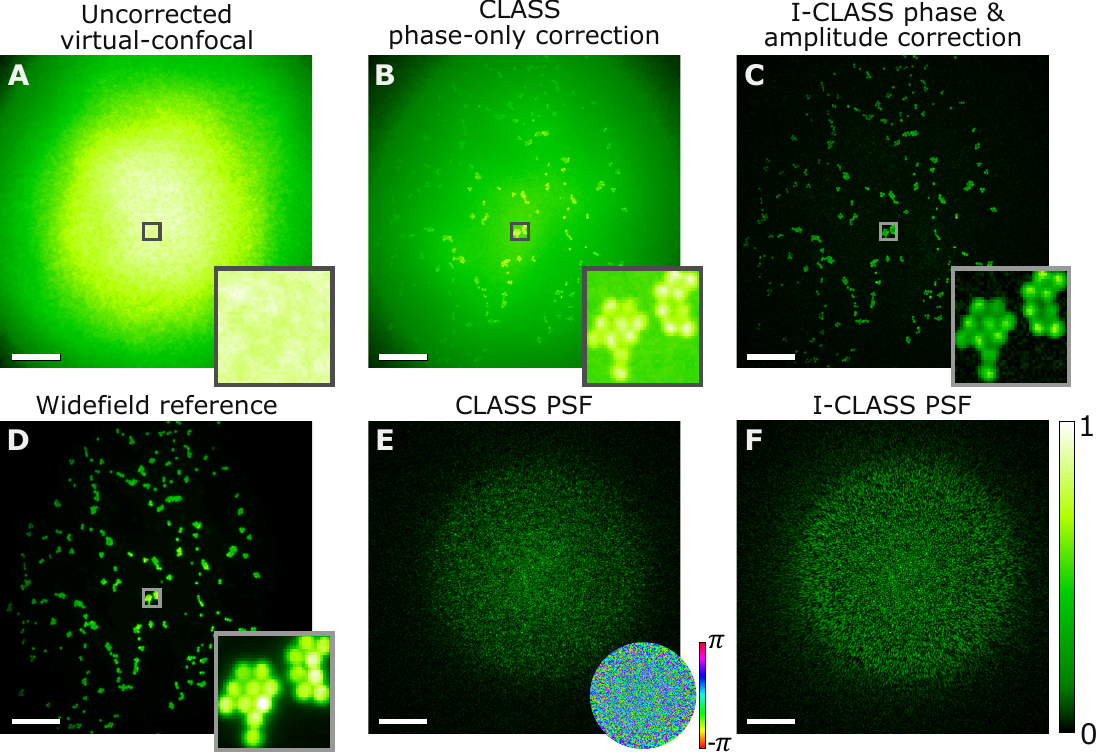}
	\caption{\textbf{Demonstration of the modified incoherent CLASS (I-CLASS) phase and amplitude corrections for isoplanatic scattering} 
 (A) Experimental uncorrected virtual-confocal image of a fluorescent beads target imaged by a microscope objective with a diffuser placed at its back focal plane.
 (B) The conventional CLASS-corrected image displays a significant haze since phase-only k-space corrections are applied. 
 (C) The modified I-CLASS phase and amplitude-corrected image provide a high-fidelity reconstruction of the ground-truth target obtained without the scattering diffuser (D). 
 (E) The correction PSF estimated by the phase-only conventional CLASS algorithm (E), inset: k-space phase correction mask. (F) same as (E) for the modified I-CLASS algorithm. Scale bars, 200 $\mu m$. Insets in (A-D) compare the marked square areas.}	\label{fig2}
\end{figure}

We demonstrate the effectiveness of I-CLASS using the experimental setup schematically depicted in Fig.~\ref{fig1}A. 
As an initial proof-of-concept experiment and to highlight the differences between the phase and amplitude correction of I-CLASS and the phase-only correction of CLASS, we consider the case of near-perfectly isoplanatic scattering. This is achieved by positioning a diffuser at the back focal plane of the microscope objective. In this configuration, the target object comprises randomly distributed fluorescent beads with an average size of 10 $\mu m$ (Fig.~\ref{fig2}).
As a starting point, we display the uncorrected 'virtual-confocal' image obtained from the diagonal of $\textbf{R}$ (Fig.~\ref{fig2}A). Explicitly, the virtual-confocal image is the standard deviation of each camera pixel and is equivalent to dynamic speckle illumination microscopy \cite{ventalon2005quasi}, or speckle-SOFI \cite{kim2015superresolution}. As expected, the image is a very low-contrast diffusive blur without apparent features, which does not resemble the target object (Fig.~\ref{fig2}D). 
Applying conventional (phase-only) CLASS correction yields a reconstruction across the entire field of view that is accompanied by a strong diffusive hazy background (Fig.~\ref{fig2}B), and an estimation of the speckled PSF from the phase-only correction (Fig.~\ref{fig2}E). 
Applying I-CLASS provides a high-contrast, high-fidelity reconstruction of the target object (Fig.~\ref{fig2}C) as imaged without the diffuser (Fig.~\ref{fig2}D), as well as an estimation of the scattering PSF. 

\begin{figure}[hbt!]
	\centering
	\includegraphics [width=0.99\textwidth,]
	{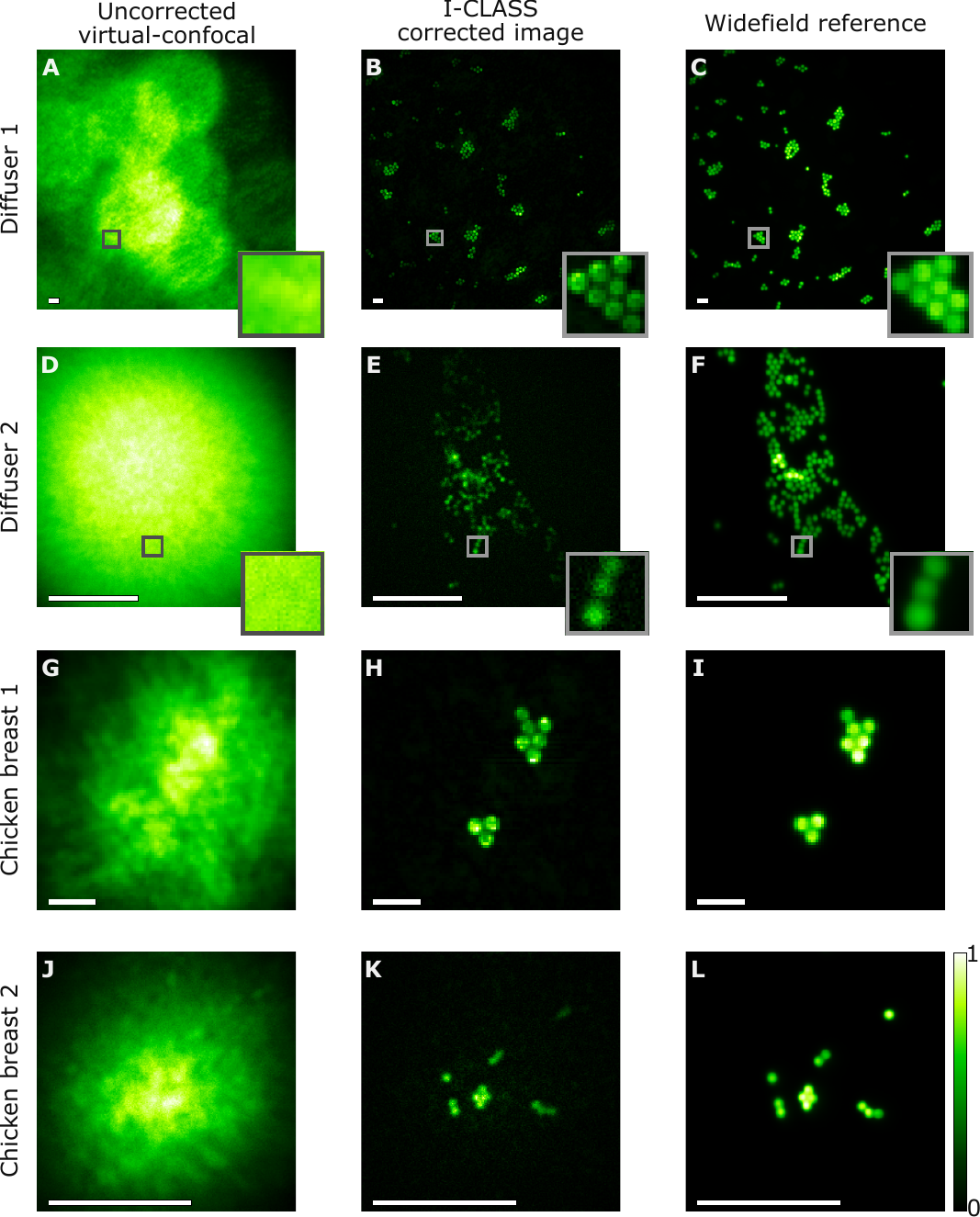}
        \caption{\textbf{Experimental I-CLASS imaging through various scattering layers}. Experimental images of fluorescent beads imaged through an optical diffuser (A-F) or a $300$-$400\mu m $ thick slices of chicken breast tissue (G-L), using two different microscope objectives: a low NA objective ($0.1$ NA, $\times 4$), and a higher NA objective ($0.6$ NA, $\times 20$). 
        Leftmost panels (A,D,G,J) display the uncorrected virtual-confocal images.
        Center column panels (B,E,H,K) display the I-CLASS corrected images, clearly reconstructing the fine details of the fluorescent bead targets, as imaged without the scattering layers (C,F,I,L).
        The scattering layers were positioned at distances of $13mm$ (A), $7mm$ (D), $5mm$ (G), and $2.5mm$ (J) from the targets. Scale bars, 30 $\mu m$}	
        \label{fig3}
        \end{figure}

To validate the technique in more realistic imaging settings, we tested its performance when various scattering layers were placed between the objective lens, as shown in Fig.\ref{fig1}A.
We studied two types of scattering layers: the same holographic diffuser used in Fig.~\ref{fig2} and a slice of $\sim 300-400\mu m$ thick chicken breast tissue.
We tested each of the scenarios using two different microscope objectives: an $\times 20$ $0.6$ NA objective and a lower magnification and lower NA objective ($\times 4$ $0.1$ NA), which allows widefield imaging of a larger field of view (FoV). 
The results of this study are presented in Fig.~\ref{fig3}: The leftmost row (Fig.~\ref{fig3}A,D,G,J) displays the uncorrected virtual-confocal images, all displaying low-contrast blurry images. The center column (Fig.~\ref{fig3}B,E,H,K) presents the I-CLASS corrected images, displaying high contrast and high-resolution images that reconstruct the fine details of the targets, composed of $2 \mu m$ diameter and $10\mu m$ diameter fluorescent beads (Fig.~\ref{fig3}C,F,I,L).

\newpage{}

\begin{figure}[h!]
	\centering
	\includegraphics [width=\textwidth,]
	{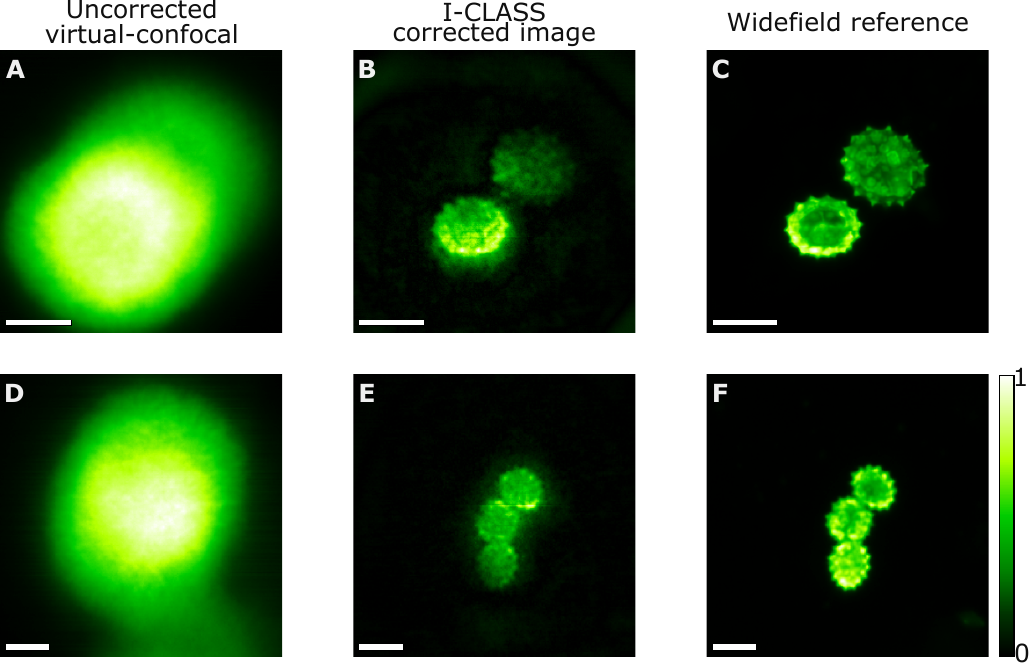}
        \caption{\textbf{Experimental I-CLASS imaging of pollen grains through a scattering diffuser}. Experimental images of pollen grains imaged through a 1$^\circ$ optical diffuser. (A-C) Tanacetum coccineum pollen grains are imaged with a $0.28$ NA, $\times 10$ objective, while (D-F) displays Dimorphotheca ecklonis pollen grains using a $0.13$ NA, $\times 4$ objective.
        Left panels (A,D) show the uncorrected virtual-confocal images.
        Center panels (B,E) display the I-CLASS corrected images, clearly reconstructing the details of the pollen grains, as imaged without the optical diffuser (C,F). The diffuser was positioned at a distance of $4mm$ (A) and $5mm$ (D) from the targets. Scale bars, 30 $\mu m$}	
        \label{fig4}
        \end{figure}

Lastly, we validate the I-CLASS reconstruction on biological specimens, specifically examining flower pollen grains. In Figure \ref{fig4}, we provide images of two flower pollens: Tanacetum coccineum (A-C) and Dimorphotheca ecklonis (D-F), both positioned behind the same $1^\circ$ holographic diffuser used in the previous experiments in Figs.~\ref{fig2}, \ref{fig3}. These samples were imaged using different objectives, $0.28$ NA $\times 10$, and $0.13$ NA $\times 4$, respectively. 
We first present the uncorrected virtual-confocal images (Fig.~\ref{fig4}A,D), characterized by their low-contrast and blurry appearance.
Subsequently, the images post I-CLASS correction (Fig.~\ref{fig4}B,E) reveal the pollen grains' details and distinct separation as imaged without the scattering layer (Fig.~\ref{fig4}C,F).

\section*{Discussion}
We have introduced and experimentally demonstrated an approach to computationally correct severe scattering distortions in a conventional widefield fluorescence microscope setup. 
As our approach is based on constructing a virtual reflection-matrix from a small number of measurements, it allows the applications of different scattering-compensation techniques developed for reflection-matrix analysis.
Similar to recently introduced techniques \cite{lee22,najar23}, our approach significantly reduces the number of required measurements by using random uncorrelated illuminations. 
In addition, our memory-efficient implementation for CTR-CLASS \cite{lee22} significantly reduces the memory requirements by avoiding the explicit calculation of the full reflection matrix, resulting in a reduction of over four orders of magnitude in memory requirement.

In contrast to recently introduced novel approaches for scattering compensation in fluorescence imaging, our approach does not require any control of the illumination \cite{daniel19,d2022physics,aizik2022fluorescent,aizik2023non} or detection paths \cite{feng23,aizik2023non,yeminy2021guidestar}, and as such, does not require any SLMs.
In addition, our approach does not rely on any prior knowledge, such as the object sparsity, and is thus not limited to localization-based image reconstruction or sparse targets \cite{boniface20,zhu22}. 
Importantly, the reflection-matrix-based scattering compensation algorithm is physically tractable and requires a small number of measurements that do not depend on the sparsity of the imaged target \cite{lee22}. 

When there is prior knowledge of the object sparsity, applying singular value decomposition (SVD)-based denoising to the virtual 'reflection matrix' can be applied, as in the coherent reflection-matrix case \cite{jo2022through,badon2016smart}. Interestingly, the SVD filtering can be applied directly on the smaller dimension measurement-matrix $\hat{\textbf{A}}$, since it shares singular values and left singular-vectors with the virtual 'reflection matrix' (Eq.~\ref{eq:two}).

We have demonstrated computational scattering correction. Physical correction of the detection and/or excitation paths using SLMs may also be possible by retrieving the coherent pupil function from the phase and amplitude I-CLASS estimated correction through a phase-retrieval step \cite{boniface20,aizik2022fluorescent}.

The main limitation for applying the I-CLASS correction is the scattering function's assumed isoplanatism (or the optical 'memory effect').
Thus, for a perfect correction, the lateral and axial extents of the target object should be smaller than $\Delta x_{mem}$ laterally, \cite{osnabrugge2017generalized}, and $\Delta z_{mem} \approx \lambda / NA^2$, axially, where $\lambda$ is the wavelength and NA is the effective numerical aperture of the detection path \cite{katz14}. 
Extending the corrected FoV beyond the memory effect would be highly valuable for many applications. When dealing with lightly-scattering media, it's possible to effectively extend the field of view for correction by separating the isoplanatic patches of the object on the image plane, achieved by cropping the image \cite{lee22}. Unfortunately, this method is unsuitable in highly scattering media where different isoplanatic patches merge within the image plane. Promising proposals have been recently introduced for anisoplanatic corrections using the coherent reflection-matrix \cite{kang2023tracing}.
In the specific context of the I-CLASS incoherent virtual reflection matrix, extending the corrected FoV axially presents a greater challenge. This is because traditional axial sectioning methods, like coherence gating, are not applicable in incoherent situations.

Finally, we note that similar to dynamic speckle illumination microscopy \cite{ventalon2005quasi} and speckle-based SOFI \cite{dertinger2009sofi, kim2015superresolution}, the lateral resolution of I-CLASS can surpass the diffraction-limited resolution of a widefield microscope by a small factor. This can be understood from the fact that I-CLASS corrects each of the $M$ fluorescent images, but instead of calculating the mean of the corrected images, the final image is their temporal standard deviation. This is numerically demonstrated on a sample that contains clear, sufficiently small features in Supplementary Fig.~S3. 

\section*{Materials and methods}
\subsection*{Numerical Simulation}

To generate the numerical results presented in Fig.~\ref{fig1}, a target object fluorescence pattern sourced from an open-access fluorescence microscopy dataset \cite{hagen2021fluorescence} was used. 
The target object pattern was multiplied by $M=150$ random intensity speckle patterns generated numerically. Each resulting frame was subsequently convoluted with a randomly generated speckled PSF. This process created $M=150$ simulated images representing the frames captured by a conventional widefield fluorescence microscope under random speckle illuminations.

\subsection*{Experimental Setup}
 
The details of the experimental setup depicted in Fig.~\ref{fig1} are as follows: The laser source is a 40mW, 488nm continuous-wave (CW) laser (Oxxius LBX-488). The random speckle illuminations are generated by rotating a holographic diffuser (RD, NEWPORT 0.5$^\circ$ + RPC Photonics EDS-1$^\circ$) placed at the excitation path $\approx$ 20cm from the microscope objective. In Fig.~\ref{fig2}, and Fig.~\ref{fig3}A-C,G-I the microscope objective is RMS4X, 0.1NA and Fig.~\ref{fig3}D-F,J-L 20X EO HR Edmund Optics, 0.6NA. 
In Fig.~\ref{fig4}A-C, the microscope objective is MY10X-803, 0.28NA and Fig.~\ref{fig4}D-F RMS4X-PF 4X, 0.13NA. 
Images are captured by an Andor Neo 5.5 sCMOS camera, integrated into a $4f$ imaging setup with a $180mm$ (Results in Fig.~\ref{fig2}, and Fig.~\ref{fig3}A-C) or $300mm$ tube lens (Results in Fig.~\ref{fig3}D-L, and Fig.~\ref{fig4}). The light is filtered with a dichroic mirror (Thorlabs DMLP490R) and emission filter (Thorlabs MF525-39). The target is made by placing fluorescent beads (Fluoresbrite YG Microspheres 10$\mu m$ for Fig.~\ref{fig2}, and Spherotech Fluorescent Yellow Particles 2$\mu m$) on a cover glass positioned at the focal plane of the objective lens. In Fig.~\ref{fig2}, Fig.~\ref{fig3}A-F and Fig.~\ref{fig4} a holographic diffuser (RPC Photonics EDC1) serves as the scattering layer, while in Fig.~\ref{fig3}G-I, and J-L the scatterer is a 300 and 400$\mu m$ thick slices of chicken breast respectively.

\subsection*{Experimental parameters}

The experimental parameters for the results displayed in Figs.~\ref{fig2}, \ref{fig3}, \ref{fig4}, including camera exposure time, image pixel count, and the scatterer's distance from the target, are outlined as follows:
In Fig.~\ref{fig2}, images are cropped to 800x800 pixels with a 9-second camera exposure, and the diffuser is positioned at the objective lens's back-focal plane.
In Fig.~\ref{fig3}A-C, the frames are first low-pass filtered from 1800x1800 to 1600x1600 pixels and are then cropped to 400x400 pixels. Each frame is captured with an exposure time of 7 seconds, with the diffuser 13mm from the object.
Fig.~\ref{fig3}D-F maintains a 400x400 pixel resolution, but the exposure time is reduced to 200 ms, with the diffuser 7mm from the object.
Fig.~\ref{fig3}G-I shows images low-pass filtered from 500x500 to 300x300 pixels, then cropped to 100x100 pixels with a 5-second exposure, and a $300\mu m$ chicken breast is positioned 5mm from the object.
Fig.~\ref{fig3}J-L feature images cropped to 250x250 pixels with a 500 ms exposure, and a $400\mu m$ chicken breast is 2.5mm from the object. 
In Fig.~\ref{fig4}A-C, images are cropped to 300x300 pixels with a 4-second exposure per frame, with the diffuser 4mm from the object.
In Fig.~\ref{fig4}D-F, the displayed images have been cropped to 200x200 pixels. Each frame is captured with an exposure time of 7 seconds, with the diffuser 5mm from the object.
In the measurements of Fig.~\ref{fig3}D-F, Fig.~\ref{fig3}J-L, and Fig.~\ref{fig4}D-F, each frame was pre-processed with Hann windowing to mitigate signal from neighboring beads and establish a zero boundary condition.

All experiments utilized $M=150$ illuminations for reconstruction, and the number of necessary illuminations is explored in supplementary S2.
The algorithm run time on a commercially available GPU (Nvidia RTX4090, 24 GB) was approximately $\sim$ 200ms per iteration for 150 camera frames at a resolution of $1400x1400$ pixels and around $\sim$ 50ms seconds per iteration for 150 camera frames at a resolution of $700x700$ pixels.

\subsection*{I-CLASS}
The full description of the I-CLASS memory-efficient, phase, and amplitude scattering compensation algorithm is fully given in Supplementary Section S1. Here, we provide a concise explanation of the I-CLASS algorithm.

\subsubsection*{Memory-efficient CLASS iterations}

The memory-efficient equivalent method for calculating the CLASS iterations using the $N \times M$ matrix $\hat{\textbf{A}}$ without the explicit computation of the $N \times N$ matrix $\textbf{R}$ can be calculated from the following formula for the t-th iteration (see Supplementary S1): 

\begin{eqnarray}
\vec{z}_{t+1}=\sum_{q=1}^{M}(\tilde{\textbf{A}}_t^* \odot ((\tilde{\textbf{A}}_t^{(ud)^*}*\tilde{\textbf{A}}_t^{(lr)})_{:,M-1} \star \tilde{\textbf{A}}_t^{(ud)}))_{:,q}
\label{eq:three}
\end{eqnarray}

\noindent where $\tilde{\textbf{A}}$ as the 2D Fourier transform of $\hat{\textbf{A}}$. Here, $\tilde{\textbf{A}}^{(ud)}$ is the matrix $\tilde{\textbf{A}}$ with each of its columns flipped upside-down, and $\tilde{\textbf{A}}^{(lr)}$ is $\tilde{\textbf{A}}$ with its rows flipped left-to-right. In addition, we denote the element-wise Hadamard product as $\odot$, the 2D convolution as $*$, and the 2D correlation operator as $\star$. $\textbf{X}^{*}$ is the element-wise complex-conjugation operator on the matrix $\textbf{X}$, and $\textbf{X}_{:,q}$ is the q-th column of $\textbf{X}$.

As in the conventional CLASS algorithm \cite{kang17}, throughout the iterative process of the I-CLASS algorithm, we solely update the k-space (OTF) phase correction with the correction:
\begin{eqnarray}
\tilde{\textbf{A}}_{t+1}=diag\{e^{i\frac{\vec{z}_t}{|\vec{z}_t|}}\}\tilde{\textbf{A}}_t
\end{eqnarray}
\noindent where the exponential and division operations are element-wise and $t=1...T$ is the iteration number. 

The phase-corrected image is then reconstructed by inverse Fourier transforming back the corrected Fourier matrix $\tilde{\textbf{A}}$ into $\hat{\textbf{A}}$ and taking the pixel-wise square root of the sum of the columns of $|\hat{\textbf{A}}|^2$ which is equivalent to the scattering-corrected virtual-confocal image (see supplementary S1), which can be written mathematically as $diag\{ \textbf{R} \} = \sum_{q=1}^{M} (\hat{\textbf{A}} \odot \hat{\textbf{A}}^*)_{:,q}$, and the CLASS corrected object $\vec{{O}}_{CLASS}$
is given by taking an element-wise square root as seen from Eq.~\ref{eq:two}.

 \subsubsection*{I-CLASS k-space amplitude correction}

 The k-space amplitude correction in the I-CLASS algorithm consists of two primary steps: (1) estimating the MTF, up to a scaling factor, using $\widehat{MTF} \equiv  \sqrt{diag\{ \tilde{\textbf{R}} \}} = \sqrt{\sum_{q=1}^{M} (\tilde{\textbf{A}} \odot \tilde{\textbf{A}}^*)_{:,q}}$ (See supplementary S1). (2) The estimated MTF is then utilized in a regularized Fourier reweighting on the last iteration CLASS corrected object $\vec{{O}}_{CLASS}$:
 
 \begin{eqnarray}
     \vec{\tilde{O}}_{I-CLASS_{i}} \equiv \frac{ \vec{\tilde{O}}_{CLASS_{i}}}{\frac{\widehat{MTF}_i}{\max_{q} \widehat{MTF}_q }+\sigma}
\end{eqnarray}

\noindent where $\sigma$ is the regularization parameter, and $\vec{\tilde{O}}_{CLASS}$ is the Fourier transformed $\vec{O}_{CLASS}$.

\subsection*{Funding}
\noindent This project was supported by the H2020 European Research Council (101002406).

\subsection*{Competing interests}
\noindent The authors declare no competing interests.

\subsection*{Code and data availability}
The code and sample datasets for the I-CLASS algorithm and measurements are available in https://github.com/Imaging-Lab-HUJI/Fluorescence-Computational-Imaging-Through-Scattering-Layers.
Additional data related to this paper may be requested from the authors.

\bibliography{bib}
\newpage
\part*{Supplementary Materials}
\author{} 
\setcounter{section}{0}
\section{Modified CLASS Algorithm (I-CLASS)}
\subsection*{Forward Model}

In the context of incoherent imaging of fluorescence targets through a disordered media, where an object is located within the optical memory effect regime \cite{feng1988correlations}, the image intensity can be described as a convolution of the intensity impulse response (PSF) with the ideal image intensity \cite{goodman2005introduction}.
For coherent illumination of a fluorescence target object, we obtain the following:

\begin{eqnarray}
{I}(\vec{r}) = (|{E}_{in}(\vec{r})*{P}^{(E)}_{ill}(\vec{r})|^2{O}(\vec{r}))*{P}(\vec{r})
\label{eq:1}.
\end{eqnarray}

Where ${P}^{(E)}_{ill}, {P}_{det}$ are the coherent illumination A-PSF and incoherent detection PSF, respectively, and ${O}(\vec{r})$ is the intensity distribution function of the object.
Thus, under coherent random illuminations, one can decompose a fluorescence measurement matrix in the $r$-basis in the following manner: ${\textbf{A}} = \textbf{P}_{det}\textbf{O}|\textbf{P}^{(E)}_{ill}\textbf{S}^{(E)}|^2$.
Notably, $\textbf{P}^{(E)}_{ill}$ and $\textbf{P}_{det}$ represent convolution matrices, the optical system input illumination coherent A-PSF and detection incoherent PSF, respectively. Furthermore, the matrix ${\textbf{O}}$ is described by a diagonal matrix, carrying the elements of ${O}(\vec{r})$ on its diagonal, and $\textbf{S}^{(E)}$ contains columns that represent the random fields in the illumination plane. For simplicity let us denote the matrix  $\textbf{S} \stackrel{\text{def}} = |\textbf{P}^{(E)}_{ill}\textbf{S}^{(E)}|^2$. Since $\textbf{S}$ is composed of independent intensity speckle-patterns originating from the same source, one can show that since the patterns are independent, we get that the correlation between a pair of pixels ${a,b}$ and speckle patterns ${i,j}$ is $\overline{I_{i_a}I_{j_b}}=\bar{I_a}\bar{I_b}+\delta_{a,b}\delta_{i,j}\bar{I_a}^2$ using $\overline{I^2}=2\bar{I}^2$, with different values due to the speckle pattern envelope \cite{goodman2007speckle}.
Leading to $\overline{(\textbf{S}\textbf{S}^{T})_{i,j}} = \sum^{M-1}_{k=0}\overline{I_{k_i}I_{k_j}} = M\bar{I_i}\bar{I_j}+M\delta_{i,j}\bar{I_i}^2$.

By using $\overline{(\textbf{S}\textbf{S}^{T})^2_{i,j}}=\sum^{M-1}_{k,k'=(0,0)}\overline{I_{k_i}I_{k_j}I_{k'_i}I_{k'_j}} = \underbrace{(M^2-M)\overline{I_{i}I_{j}}^2}_{k\neq k'}+M\overline{I^2_{i}I^2_{j}}$ 
and $\overline{I^n_i}=n!\overline{I_i}^n$, one can show that:$\frac{\Delta(\textbf{S}\textbf{S}^{T})_{i,j}}{(\textbf{S}\textbf{S}^{T})_{i,j}} \sim \frac{1}{\sqrt{M}}$
($\Delta X$ denotes the standard deviation of $X$). 

Thus, we conclude that to a good approximation $\textbf{S}\textbf{S}^{T} \approx Mdiag(\bar{I}\odot \bar{I}) + M\vec{\bar{I}}\vec{\bar{I}}^T$ where $\vec{\bar{I}}$ is the vector that holds the mean intensity values for each pixel (and $\odot$ is the element-wise Hadamard product).
Hence, by implementing pixel-wise mean reduction, we define $\hat{\textbf{S}}\stackrel{\text{def}} = \textbf{S} (\mathbf{I}-\frac{1}{M}\cdot\vec{1}\vec{1}^T)$,where $\vec{1}$ represents a column vector comprising entirely of ones ($\forall i\in[N] :
\vec{1}_{i}=1$). Importantly, this transformation renders $\hat{\textbf{S}}$ uncorrelated:

\begin{multline}
\hat{\textbf{S}} \hat{\textbf{S}}^T = \textbf{S}(\mathbf{I}-\frac{1}{M}\cdot\vec{1}\vec{1}^T)^2\textbf{S}^T = \textbf{S}(\mathbf{I}-\frac{1}{M}\cdot\vec{1}\vec{1}^T)\textbf{S}^T = 
\\
\textbf{S}\textbf{S}^T-\frac{1}{M}
\underbrace{(\textbf{S}\vec{1})}_{\approx M\vec{\bar{I}}}
\underbrace{(\textbf{S}\vec{1})^T}_{\approx M\vec{\bar{I}}^T}\approx Mdiag(\bar{I}\odot \bar{I}) + M\vec{\bar{I}}\vec{\bar{I}}^T- M\vec{\bar{I}}\vec{\bar{I}}^T = Mdiag(\bar{I}\odot \bar{I})
\label{eq:6}
\end{multline}

leading to the approximation ${\hat{\textbf{S}}} \hat{\textbf{S}}^T \approx \textbf{D}$, for $\textbf{D}\stackrel{\text{def}} =  Mdiag(\bar{I}\odot \bar{I})$.
Notably, from a mathematical perspective, it becomes apparent that by subtracting the temporal mean from the measurement matrix $\textbf{A}$, we effectively replicate an illumination pattern with uncorrelated illuminations $\hat{\textbf{S}}$:

\begin{eqnarray}
\hat{\textbf{A}}\stackrel{\text{def}} = {\textbf{A}}(\mathbf{I}-\frac{1}{M}\cdot\vec{1}\vec{1}^T) = {\textbf{P}_{det}}{\textbf{O}}{\textbf{S}}(\mathbf{I}-\frac{1}{M}\cdot\vec{1}\vec{1}^T)
 = {\textbf{P}_{det}}{\textbf{O}}\hat{\textbf{S}}
 \label{eq:7}
\end{eqnarray}
and from the calculation of the covariance matrix, we overall get:
\begin{eqnarray}
\textbf{R}=\hat{\textbf{A}}\hat{\textbf{A}}^T = {\textbf{P}_{det}}\textbf{O}\hat{\textbf{S}}\hat{\textbf{S}}^T\textbf{O}\textbf{P}_{det}^T \approx \textbf{P}_{det}{\textbf{O}_{eff}} \textbf{P}_{det}^T
\label{eq:8}
\end{eqnarray}
for $\textbf{O}_{eff}\stackrel{\text{def}}=\textbf{O}^2 \textbf{D}$, an incoherent fluroesecent analog to a virtual reflection matrix.

Utilizing the multiplication-convolution relationship in k-space, we can decompose the Fourier transform of the reflection matrix into: 
\begin{align}
\tilde{\textbf{R}}\equiv\mathcal{F}\textbf{R}\mathcal{F}^{\dagger} \approx  \mathcal{F}\textbf{P}_{det}{\textbf{O}_{eff}} \textbf{P}_{det}^T\mathcal{F}^{\dagger} = && \label{eq:18} \\ \nonumber 
\mathcal{F}\textbf{P}_{det}\mathcal{F}^{\dagger}\mathcal{F}\textbf{O}_{eff}\mathcal{F}^{\dagger}(\mathcal{F}\textbf{P}_{det}\mathcal{F}^{\dagger})^{\dagger}={\tilde{\textbf{P}}_{det}}{\tilde{\textbf{O}}_{eff}}{\tilde{\textbf{P}}^{\dagger}_{det}} 
\end{align}
(where $\mathcal{F}$ represents the unitary DFT matrix).

In this Fourier domain representation, $\tilde{\textbf{P}}_{det}$ which is the Fourier transform of $\textbf{P}_{det}$, takes on a diagonal form, reflecting the fact that the convolution operations in real space become simple multiplications in Fourier space. In the case of a unitary phase-only distortion (which is the model in the coherent case), these diagonal matrices contain the random phases $e^{i\phi_{det}(\vec{k})}$ and $e^{-i\phi_{det}(\vec{k})}$, respectively, which the CLASS algorithm aims to correct. On the other hand, the matrix $\tilde{\textbf{O}}_{eff}$ which is the Fourier transform of $\textbf{O}_{eff}$, now adapts the form of a Toeplitz convolution matrix, with the Fourier components of the object.

\subsection*{CLASS algorithm}

The CLASS algorithm \cite{kang17} aims to correct the distortion, $\tilde{\textbf{P}}_{det}$, and retrieve the ideal image intensity ${\tilde{\textbf{O}}_{eff}}$. Mathematically, this requires finding the inverse of $\tilde{\textbf{P}}_{det}$ which, being modeled as a unitary transformation with only phase aberrations, is equivalent to finding $\tilde{\textbf{P}}_{det}^{\dagger}$.

The standard CLASS algorithm, handling a full system reflection matrix in the form of  $\textbf{R}={\tilde{\textbf{P}}_{det}}{\tilde{\textbf{O}}}{\tilde{\textbf{P}}_{ill}}$, addresses a single distortion during each iteration, alternately handling $\tilde{\textbf{P}}_{det}$ and $\tilde{\textbf{P}}_{ill}$. However, in CTR-CLASS \cite{lee22} where $\tilde{\textbf{P}}_{ill} = \tilde{\textbf{P}}_{det}$, exclusively tackles the "right" matrix $\tilde{\textbf{P}}_{det}$ in each iteration and fixes the matrix from both sides.

Given that we model the matrix $\tilde{\textbf{R}}$ as the product of a Toeplitz matrix $\tilde{\textbf{O}}_{eff}$ and a diagonal matrix $\tilde{\textbf{P}}_{det}$, it follows that if we displace each column proportionally to its index, the resultant matrix should exhibit columns of equal values which contain the spectrum of $\tilde{O}_{eff}(\vec{k})$, albeit with distinct global phases for each column. Consequently, if we compute the average of these columns, we obtain a reasonably accurate estimation of $\tilde{O}_{eff}(\vec{k})$, given that the phase averages out.\\
Now that we possess an approximation of $\tilde{O}_{eff}(\vec{k})$, we can determine the overall phase for each column. This can be achieved by calculating the correlation angle between the j-th column and the estimated $\hat{\tilde{O}}_{eff}(\vec{k})$:

\begin{eqnarray}
\hat{\phi}_{det}({k_{j})} = arg\{<\hat{\tilde{O}}_{eff}^*(\vec{k})\tilde{O}_{eff}(\vec{k})e^{i\phi_{det}(k_{j})}>_k\}
\label{eq:9}
\end{eqnarray}
by incorporating the terms $e^{-i\hat{\phi}_{det}(\vec{k})}$ into a diagonal matrix $\hat{\tilde{\textbf{P}}}_{det}$ we can fix $\tilde{\textbf{R}}$ by $\tilde{\textbf{R}}_n=\hat{\tilde{\textbf{P}}}_{det}^{\dagger}\tilde{\textbf{R}}_{n-1}\hat{\tilde{\textbf{P}}}_{det}$. This iterative approach consistently leads to the convergence of the correct phase correction.
We now write the iteration in a matrix notation, by defining:
\begin{eqnarray}
\vec{z}\stackrel{\text{def}} =  \tilde{\textbf{R}}\tilde{\textbf{R}}^{T}_{n_s}(\tilde{\textbf{R}}_{n_s}\vec{1})^*=(\tilde{\textbf{R}}^{\dagger}_{n_s}\tilde{\textbf{R}}_{n_s}\vec{1})^*=\vec{1}^T\tilde{\textbf{R}}^{\dagger}_{n_s}\tilde{\textbf{R}}_{n_s}
\label{eq:10}
\end{eqnarray}

With this notation, we obtain:  $\hat{\phi}_{det}({k_{j})}=\frac{\vec{z}_j}{|\vec{z}_j|}$

Here, $\tilde{\textbf{R}}_{n_s}$ represents the shifted version of $\tilde{\textbf{R}}^{(ud)}$, which can be expressed as:
\begin{eqnarray}
\forall i\in[2N-1] \forall j\in[N]:
\tilde{\textbf{R}}_{n_{s_{i,j}}}=\sum^{N-1}_{a=0}\sum^{N-1}_{b=0}S_{i,j,a,b}\tilde{\textbf{R}}^{(ud)}_{n_{a,b}}
\label{eq:11}
\end{eqnarray}
using $S_{i,j,a,b} = \delta_{j,b}\delta_{i,a+b}$ . Additionally, we use $\tilde{\textbf{R}}^{(ud)}_{n_{i,j}} = \tilde{\textbf{R}}_{N-1-i,j}$ i.e. flipping each column of $\tilde{\textbf{R}}$, a measure that was taken for the sake of index convenience.
(By denoting $\textbf{R}^{(ud)}_{{i,j}} = \textbf{R}_{N-1-i,j}$ we can get $\tilde{\textbf{R}}^{(ud)} = \tilde{\textbf{A}}^{(ud)}\tilde{\textbf{A}}^\dagger$)

\subsection*{Memory Complexity}
Given that we capture only $M$ images from distinct random illuminations, each containing $N$ pixels, where $M<<N$, it is advantageous to avoid explicitly constructing the matrix $\textbf{R}$ as ${\hat{\textbf{A}}}{\hat{\textbf{A}}}^{\dagger}$, an $N$x$N$, matrix, due to its high memory demand. All the necessary information should be encapsulated within $\textbf{R}$, which is of size $N$x$M$. As a result, we show an equivalent mathematical expression for a CLASS iteration using $\hat{\textbf{A}}$ without directly computing $\textbf{R}$.

We note that since the DFT matrix \( \mathcal{F} \) is unitary, we can apply a two-dimensional Fourier transform to the measured images before including them in \( {\hat{\textbf{A}}} \), thus obtaining \( \tilde{\textbf{R}} = \mathcal{F}\textbf{R}\mathcal{F}^{\dagger} = \mathcal{F}{\hat{\textbf{A}}}{\hat{\textbf{A}}}^{\dagger}\mathcal{F}^{\dagger} = \mathcal{F}{\hat{\textbf{A}}}\mathcal{F}\mathcal{F}^{\dagger}{\hat{\textbf{A}}}^{\dagger}\mathcal{F}^{\dagger} \). Therefore, we can concentrate on $\tilde{\textbf{A}}=\mathcal{F}{\hat{\textbf{A}}}\mathcal{F}^{\dagger}$.

First, we find the vector elements of $\vec{z} = \vec{1}^T\tilde{\textbf{R}}^{\dagger}_{n_s}\tilde{\textbf{R}}_{n_s}$ (Eq. \ref{eq:10}):

\begin{multline}
z_j = \sum^{N-1}_{i=0}(\tilde{\textbf{R}}^{\dagger}_{s}\tilde{\textbf{R}}_{s})_{i,j} = \sum^{N-1}_{i=0}\sum^{2N-2}_{m=0}\sum^{N-1}_{k=0}\sum^{N-1}_{l=0}\sum^{N-1}_{a=0}\sum^{N-1}_{b=0}S_{m,j,a,b}\tilde{\textbf{R}}^{(ud)}_{a,b}S_{m,i,k,l}\tilde{\textbf{R}}^{(ud)^*}_{k,l} = \\ \sum^{N-1}_{i=0}\sum^{N-1}_{k=0}\sum^{N-1}_{l=0}\sum^{N-1}_{a=0}\sum^{N-1}_{b=0}\tilde{\textbf{R}}^{(ud)^*}_{k,l}\tilde{\textbf{R}}^{(ud)}_{a,b}\sum^{2N-2}_{m=0}S_{m,j,a,b}S_{m,i,k,l}= \\\sum^{N-1}_{i=0}\sum^{N-1}_{k=0}\sum^{N-1}_{l=0}\sum^{N-1}_{a=0}\sum^{N-1}_{b=0}\tilde{\textbf{R}}^{(ud)^*}_{k,l}\tilde{\textbf{R}}^{(ud)}_{a,b}\delta_{j,b}\delta_{i,l}\sum^{2N-2}_{m=0}\delta_{m,a+b}\delta_{m,k+l}= \\ (2N-1)\sum^{N-1}_{i=0}\sum^{N-1}_{k=0}\sum^{N-1}_{a=0}\tilde{\textbf{R}}^{(ud)}_{a,j}\tilde{\textbf{R}}^{(ud)^*}_{k,i}\delta_{a+j,k+i}\propto\\\sum^{N-1}_{i=0}\sum^{N-1}_{k=0}\sum^{N-1}_{a=0}\sum^{M-1}_{n=0}\sum^{M-1}_{m=0}\tilde{\textbf{A}}^{(ud)}_{a,n}\tilde{\textbf{A}}^*_{j,n}\tilde{\textbf{A}}^{(ud)^*}_{k,m}\tilde{\textbf{A}}_{i,m}\delta_{a+j,k+i} = \\\sum^{N-1}_{k=0}\sum^{N-1}_{a=0}\sum^{M-1}_{n=0}\tilde{\textbf{A}}^{(ud)}_{a,n}\tilde{\textbf{A}}^*_{j,n}\sum^{M-1}_{m=0}\tilde{\textbf{A}}^{(ud)^*}_{k,m}\tilde{\textbf{A}}_{a+j-k,m}\sum^{N-1}_{i=0}\delta_{a+j,k+i}
 = \\\sum^{N-1}_{a=0}\langle \vec{\tilde{\textbf{A}}}_{j,:},\vec{\tilde{\textbf{A}}}^{(ud)}_{a,:} \rangle\sum^{min(N-1,a+j)}_{k=max(0,a+j-(N-1))}\langle \vec{\tilde{\textbf{A}}}^{(ud)}_{k,:},\vec{\tilde{\textbf{A}}}_{a+j-k,:} \rangle \label{eq:12}
\end{multline}

where we denote $\vec{\textbf{A}}_{c,:}$ as the $c-th$ column of A. By denoting ${\textbf{A}}^{(lr)}_{a,b}={\textbf{A}}_{a,M-1-b}$ and noticing that using full sized (using zero-padding) convolution we get:
\begin{eqnarray}
({\textbf{A}}^{(ud)^*}*{\textbf{A}}^{(lr)})_{a+j,M-1} = \sum^{min(N-1,a+j)}_{k=max(0,a+j-(N-1))}\langle {\textbf{A}}^{(ud)}_{k,:},{\textbf{A}}_{a+j-k,:} \rangle \label{eq:13}
\end{eqnarray}
Plugging Eq.\ref{eq:13} into Eq.\ref{eq:12}, we get:

\begin{eqnarray}
\vec{z}=\sum^{N-1}_{a=0}(\tilde{\textbf{A}}^*\vec{\tilde{\textbf{A}}}_{a,:})\odot(\tilde{\textbf{A}}^{(ud)^*}*\tilde{\textbf{A}}^{(lr)})_{a:a+N,M-1}
\label{eq:14}
\end{eqnarray}
Where $\odot$ denotes the Hadamard Product (element-wise multiplication) between two vectors. And the phase correction for each iteration is received by $\hat{\vec{\phi}} = \frac{\vec{z}}{|\vec{z}|}$ (element-wise division) .

We can also remove the sum over $a$. Lets start by denoting $\vec{b}\stackrel{\text{def}} = (\tilde{\textbf{A}}^{(ud)^*}*\tilde{\textbf{A}}^{(lr)})_{:,M-1}$.
Writing again the elements of $\vec{z}$:

\begin{equation}
z_i=(\sum^{N-1}_{a=0}(\tilde{\textbf{A}}^*\vec{\tilde{\textbf{A}}}_{a,:})\odot\vec{b}_{a:a+N})_i =\sum^{M-1}_{j=0} \tilde{\textbf{A}}^*_{i,j}\sum^{N-1}_{a=0}\tilde{\textbf{A}}^{(ud)}_{a,j}\vec{b}_{a+i}=\sum^{M-1}_{j=0} \tilde{\textbf{A}}^*_{i,j}(\vec{b} \star \tilde{\textbf{A}}^{(ud)})_{i,j}
\label{eq:15}
\end{equation}
where $\star$ is a 2D cross-correlation.
Finally, the correction can get written simply as:
\begin{equation}
\vec{z}_{t+1}=(\tilde{\textbf{A}}_t^* \odot ((\tilde{\textbf{A}}_t^{(ud)^*}*\tilde{\textbf{A}}_t^{(lr)})_{:,M-1} \star \tilde{\textbf{A}}_t^{(ud)}))\vec{1}
\label{eq:16}
\end{equation}
therefore, after each iteration $t = 1...T$, we get 
$\tilde{\textbf{A}}_{t+1}=diag\{e^{i\frac{\vec{z}_{t+1}}{|\vec{z}_{t+1}|}}\}\tilde{\textbf{A}}_t$, where the exponential and division operations are element-wise. 
Since building this correction diagonal matrix is memory inefficient, we correct each q-th column of $\tilde{\textbf{A}}_t$ with the element-wise multiplication: $ e^{i\frac{\vec{z}_{t+1}}{|\vec{z}_{t+1}|}} \odot \tilde{\textbf{A}}_{t_{:,q}}$.

After all iterations, to restore the object approximation $\hat{O}_{eff}=diag\{\textbf{R}\}$ we then calculate:
\begin{eqnarray}
\hat{O}_{eff_i}=\textbf{R}_{i,i}=\mathcal{F}^{\dagger}\tilde{\textbf{R}}\mathcal{F}_{i,i}=\mathcal{F}^{\dagger}\tilde{\textbf{A}}\tilde{\textbf{A}}^\dagger\mathcal{F}_{i,i}=\mathcal{F}^{\dagger}\tilde{\textbf{A}}\mathcal{F}\mathcal{F}^{\dagger}\tilde{\textbf{A}}^\dagger\mathcal{F}_{i,i}=\sum^{N-1}_{m=0}|\mathcal{F}^{\dagger}\tilde{\textbf{A}}\mathcal{F}|^2_{i,m}
\label{eq:17}
\end{eqnarray}
Overall, the object can be written as: $\vec{\tilde{O}}_{eff} =|iFFT2(\tilde{\textbf{A}})|^2\vec{1}$, which is the sum over the columns of $|{\hat{\textbf{A}}}|^2$.
We conclude that the improvement in the memory-efficient I-CLASS algorithm now has memory complexity $O(MN)$ instead of $O(N^2)$, an improvement of $N/M$, which in our experiments is $\sim$ $10^4$.

\subsection*{Regularized Fourier Reweighting}

In the incoherent imaging scenario mentioned in the main text, non-unitary distortion and amplitude modulation cause haze in the phase-corrected image. Thus, in estimating the Modulation Transfer Function (MTF), the absolute value of the Optical Transfer Function (OTF) is essential. 

In the preceding discussion, and as demonstrated in Eq.~\ref{eq:18}, we observe that ${\tilde{\textbf{P}}_{det}}$ is structured as a diagonal matrix containing on its diagonal  ${\tilde{\textbf{P}}_{det_{i,i}}}=MTF(k_i)e^{i\phi_{det}(k_i)}$, while ${\tilde{\textbf{O}}_{eff}}$ takes the form of a convolution matrix. This allows us to express the elements of $\tilde{\textbf{R}}$ as follows:

\begin{eqnarray}
\tilde{\textbf{R}}_{i,j} = \sum_{a,b}{\tilde{\textbf{P}}_{det_{i,a}}} \tilde{\textbf{O}}_{eff_{a,b}}{\tilde{\textbf{P}}^{\dagger}_{det_{b,j}}}=\sum_{a,b}{\tilde{\textbf{P}}_{det_{i,i}}} \delta_{i,a} \vec{\tilde{O}}_{eff_{a-b}}{\tilde{\textbf{P}}^*_{det_{j,j}}} \delta_{b,j}
\label{eq:19}
\end{eqnarray}
this formulation leads to the following expression for the diagonal elements:

\begin{eqnarray}
\tilde{\textbf{R}}_{i,i} = \vec{\tilde{O}}(0) {\tilde{\textbf{P}}_{det_{i,i}}}{\tilde{\textbf{P}}^*_{det_{i,i}}} \propto |{\tilde{\textbf{P}}_{det_{i,i}}}|^2
\label{eq:20}
\end{eqnarray}
Consequently, this relationship enables the estimation of the MTF up to a scaling factor, by:
$\widehat{MTF} \equiv \sqrt{diag\{\tilde{\textbf{R}}\}}$.
Similarly to the approach outlined in Eq. \ref{eq:17}, the MTF can be directly computed from $\tilde{\textbf{A}}$, bypassing the need for explicit calculation of $\tilde{\textbf{R}}$. This is achieved by summing over the columns of $|{\tilde{\textbf{A}}}|^2$.

Overall, the k-space amplitude correction is estimated by taking the amplitude correction $\widehat{MTF}$ as input for a regularized Fourier reweighting process, where each Fourier component of $\vec{{O}}_{CLASS}$, denoted as  $\vec{\tilde{O}}_{CLASS}$, is divided by a factor given by:
\begin{eqnarray}
     \vec{\tilde{O}}_{I-CLASS_{i}} \equiv \frac{ \vec{\tilde{O}}_{CLASS_{i}}}{\frac{\widehat{MTF}_i}{\max_{q} \widehat{MTF}_q }+\sigma}
\end{eqnarray}
\noindent where $\sigma$ is the regularization parameter.

Although one can utilize various deconvolution methods, such as Wiener or Richardson-Lucy, our empirical observations have consistently shown that regularized deconvolution yields superior results.

\begin{figure}[ht!]
	\centering
	\includegraphics [width=\textwidth,]
	{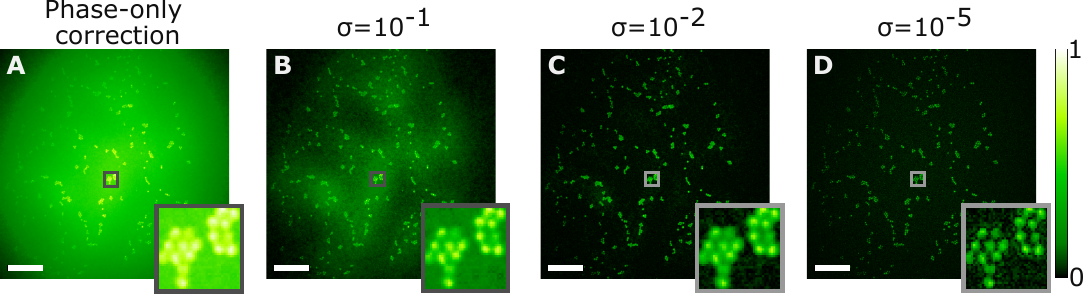}
        \renewcommand{\thefigure}{S1}
    \caption{\textbf{Regularized Fourier-reweighting with varying regularization parameter}. (A) Corrected confocal image via phase-only OTF correction shows significant hazing due to uncorrected amplitude attenuation. (B) The result of a too-high regularization parameter resulted in the inclusion of haze. (C) Shows the scenario where the regularization parameter optimally reduces noise while preserving high-frequency details. (D) Illustrates the effect when the regularization parameter (sigma) is set too low, including noisy high-frequency bands. }\label{fig1_supp}
    \end{figure}

The impact of different regularization parameters is showcased through I-CLASS corrections in Fig.~\ref{fig1_supp}. Fig.~\ref{fig1_supp}A features the CLASS correction, similar to taking the regularization parameter $\sigma $ to $\infty$. Fig.\ref{fig1_supp}B displays the result of an excessively high regularization parameter, leading to haze in the image. Fig.\ref{fig1_supp}C demonstrates an optimal balance between noise reduction and the preservation of high-frequency details. Lastly, Fig.~\ref{fig1_supp}D presents the effect of setting the regularization parameter too low, resulting in the inclusion of noisy high-frequency bands.

In summary, the I-CLASS algorithm proceeds as follows: initially, it employs the memory-efficient version of the CTR-CLASS algorithm, followed by estimating the MTF from the diagonal of the Fourier-transformed reflection matrix and finally applying deconvolution with the estimated MTF.

\newpage

\section{Number of illuminations required for image reconstruction}

To assess the effect of random illuminations on reconstruction quality, we performed a series of reconstructions of the measurements presented in Fig.2 of the main text, using various numbers of illumination subsets. The results are given in Supplementary Fig.~\ref{fig2_supp}, where we display the reconstruction outcomes for subsets from 80 to 150 images. The effectiveness of our reconstruction technique is evaluated by comparing the maximum normalized cross-correlation with the virtual confocal measurement without the scattering layer (Fig.~\ref{fig2_supp}A). Consistent with prior CTR-CLASS reconstructions \cite{lee22}, it is observed that when the compression ratio (CR) is above some threshold, neither the image contrast nor the spatial resolution of the reconstructed image was significantly diminished by the reduction of CR (Fig.~\ref{fig2_supp}B-C). And, when the reconstructions fall below a certain threshold (Fig.~\ref{fig2_supp}D), the reconstruction process struggles to accurately identify the correct aberration map, leading to an object-image that resembles a more focused image, as previously shown in \cite{lee22}.

\begin{figure}[htb!]
	\centering
	\includegraphics [width=0.98\textwidth,]
	{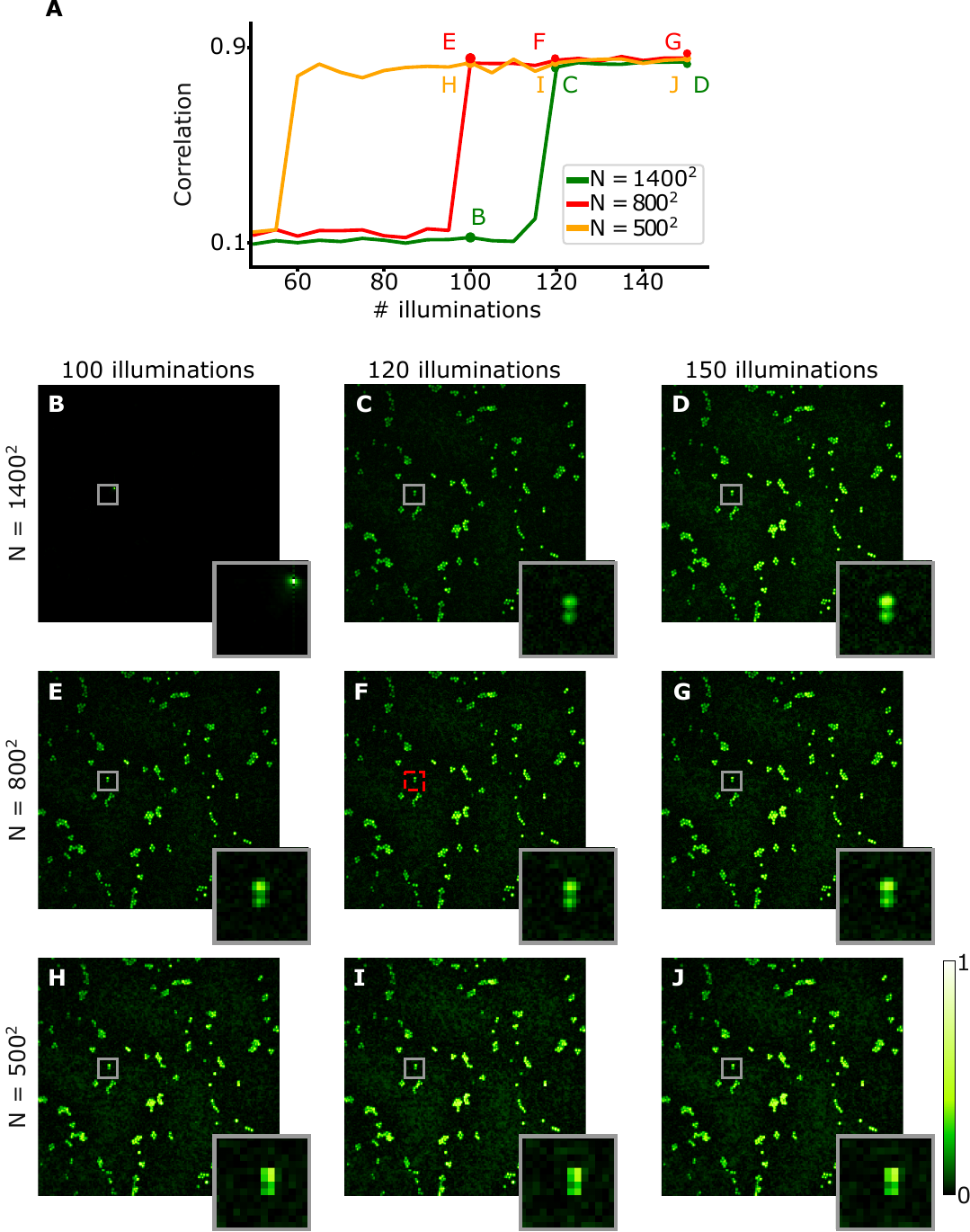}
        \renewcommand{\thefigure}{S2}
    \caption{\textbf{Investigation of required illuminations for image reconstruction}: (a) Maximum normalized cross-correlation of reconstructions from Fig.~2, examining its variation with different frame counts used for reconstruction. Three scenarios with varied input pixel numbers are presented: (B-D) Without low-pass filtering (green line), (E-G) Low-pass filtering from 1400x1400 to 800x800 pixels (red line), and (H-J) Low-pass filtering from 1400x1400 to 500x500 pixels (orange line). Each section displays reconstructions using 100, 120, and 150 frame counts.}\label{fig2_supp}
    \end{figure} 

\newpage

\section{Resolution improvement of I-CLASS}

As mentioned in the main text, our method's lateral resolution can exceed that of a conventional widefield fluorescence microscope's diffraction-limited resolution. This is numerically demonstrated in Supplementary Fig.\ref{fig3_supp}. This improvement in resolution is analogous to advancements seen in dynamic speckle illumination microscopy \cite{ventalon2005quasi} and speckle-based SOFI \cite{dertinger2009sofi, kim2015superresolution}. In Fig.\ref{fig3_supp}, we clearly illustrate this improvement. In Fig.\ref{fig3_supp}A, we present an uncorrected virtual-confocal image of a simulated USAF resolution test target. Fig.~\ref{fig3_supp}B, displays the I-CLASS corrected image. A zoomed-in section of element 1, with a vertical cross-section, highlights the improvement in resolution, which cannot be resolved at the widefield reference image shown in Fig.~\ref{fig3_supp}C.

\begin{figure}[htb!]
	\centering
	\includegraphics [width=0.98\textwidth,]
	{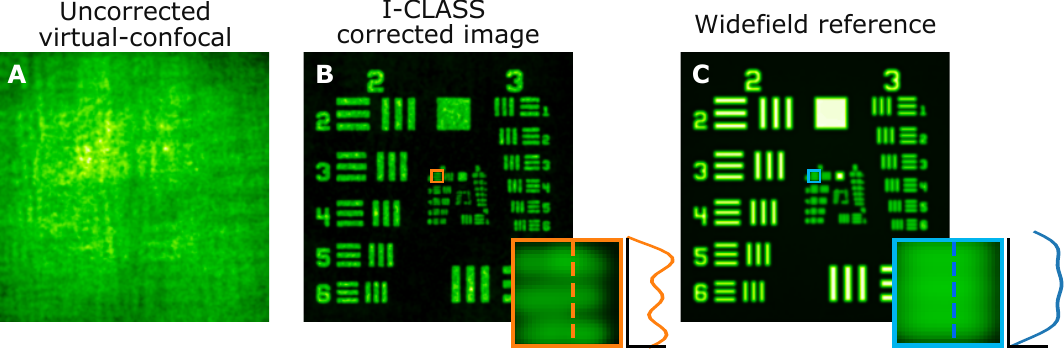}
        \renewcommand{\thefigure}{S3}
        \caption{\textbf{Numerical resolution improvement of I-CLASS}: (A) Uncorrected virtual-confocal image of a simulated USAF resolution test target. (B) I-CLASS corrected image, with a focused zoom-in on a cross-section showcasing the improved resolution capable of resolving elements in group 1. This level of detail contrasts with the widefield reference image in (C), which fails to resolve the same elements.}  \label{fig3_supp}
        \end{figure} 
        
\end{document}